\documentstyle[preprint,aps]{revtex}
\tighten
\draft
\begin{document}
\title {Noise in an {\rm ac} biased junction.\\
Non-stationary Aharonov-Bohm effect.}
\author{G.B. Lesovik\cite{*}}
\address
{Universit{\" a}t zu K{\" o}ln,\\
Institut f{\" u}r Theoretische Physik,\\
Z{\" u}lpicher Str. 77, D-5000 K{\" o}ln 41, Germany}
\author{L.S.Levitov\cite{**}}
\address{Massachusetts Institute of Technology,\\
12-112, Department of Physics,\\
77 Massachusetts Ave., Cambridge, MA 02139}
\maketitle

\begin{abstract}
We study excess noise in a quantum  conductor  in  the
presence of constant voltage and alternating external field. Due
to a two  particle interference effect caused by Fermi
correlations the noise is sensitive  to  the  phase  of  the  time
dependent  transmission  amplitude.  We compute spectral density
and show that at $T=0$ the noise has singular dependence on  the
{\rm  dc}  voltage  $V$ and the {\rm ac} frequency $\Omega$ with
cusplike  singularities  at  integer  $eV/\hbar\Omega$.  For   a
metallic  loop  with  an  alternating  flux the phase sensitivity
leads to an oscillating dependence of
the strengths of the cusps on the flux  amplitude.

\end{abstract}
\narrowtext

There is  a  variety  of  phenomena  related  with  the  quantum
coherence   of   transport  in  small  conductors\cite{1}:  weak
localization, Aharonov-Bohm effect with the flux quantum  $hc/2e$,
universal  conductance  fluctuations, etc. Each of these
effects can also be seen in the spectrum of  noise,  equilibrium
or non-equilibrium. The equilibrium noise is simply proportional
to conductance according to the fluctuation-dissipation theorem.
The  non-equilibrium  noise  in coherent conductors is expressed
through eigenvalues of  the  scattering  matrix\cite{2,3,4},  and,
therefore,  is  also  related  with the conductance, though in a
less trivial way. For that reason all  the  coherence  phenomena
are present in the non-equilibrium noise as well. However, for a
better  understanding  of  transport  in  small conductors it is
interesting to analyze the converse line of thinking and to look
for coherence effects that are present  in  the  noise  but  are
absent  in  the  conductance.  Such  effects, if they exist, are
genuinely many-particle\cite{5}, otherwise they would show up in
the conductance.
As  long as we are
talking about non-interacting fermions  it  is  only  statistics
that  can produce such coherence. A purpose of this letter is to
describe  an  effect   caused   by   two-particle
statistical  correlations  that  leads to phase sensitivity of a
two-particle observable, i.e., of electric noise, but  does  not
affect  one-particle  observables,  e.g., conductance. The phase
sensitivity manifests itself in an oscillating dependence on the
amplitude of an {\rm ac} flux, in many aspects  similar  to  the
A-B  effect. However, it will occur in a single-connected conducting
loop, i.e., in the geometry where the normal A-B effect is absent.

Let us specify which coherence effects we are going to study.
In simple words, when an electron is scattered inside a
conductor its wavepacket splits into two portions, forward and
backward, presenting a choice to the electron to be either
transmitted or reflected with the probabilities $D$ and $1-D$.
Part of this picture of the wavepacket splitting, involving the
relation of $D$ with the conductance\cite{6,7} and of $D(1-D)$ with the
noise\cite{2,3,4,8} is well understood. However, there is another
part, quite unusual, related with the behavior of current
fluctuations in the time domain. Recently, we studied the
distribution of the charge transmitted through a resistor during
fixed interval of time\cite{9}. We found that the distribution is
very close to the binomial, which means that the attempts to
have electron transmitted are highly correlated in time. (Were
the sequence of the attempts perfectly periodic the distribution
would be exactly binomial.) The origin of the correlation is the Pauli
principle that forbids passing of electrons through the resistor
simultaneously. The attempts follow almost periodically, spaced
by the interval $h/eV$. Because the periodicity is not perfect
it does not affect the average current, but shows up in its second
moment, i.e., noise, leading\cite{10} at zero temperature to a
sharp edge of the spectral density of excess noise $S_\omega$
near $\omega_0= eV/\hbar$: $S_\omega = 2{e^2\over\pi
}D(1-D)\hbar (\omega_0 - |\omega |)$ for $|\omega | < \omega_0$,
$0$ otherwise. (Excess noise is the difference of the actual
noise and the equilibrium $S_\omega=2{e^2\over\pi }D\hbar\omega\
{\rm coth}(\hbar\omega/2T)$.) The corresponding current-current
correlation in time is
${\langle}{\langle}j(t)j(t+{\tau}){\rangle}{\rangle}= {2e^2
\over {\pi}^2}D(1-D) {\rm sin}^2({\omega}_0{\tau}/2)/{\tau}^2$,
oscillating with the period $2\pi/\omega_0$ and decaying.

Having realized that the frequency $eV/\hbar$ is  characteristic
for  the  time correlation of the attempts one has to think of a
simple  experimental  situation  were  the  presence   of   this
frequency  could  be studied. It is natural to consider a system
driven both {\rm dc} and {\rm ac}, and to look for  the  effects
of  commensurability  of $\Omega$ and $eV/\hbar$, where $\Omega$
is the frequency of the {\rm ac} bias and $V$ is  the  {\rm  dc}
voltage.  In  this letter we study such a system and demonstrate
that due to  the  {\rm  ac}  bias  the  singularity  at  $\omega
=\omega_0$  can be shifted down to zero frequency thus making it
easier to observe. Below,  we  compute  the  noise  in  a  model
resistor  in the presence of combined {\rm dc}-{\rm ac} bias and
find that the low frequency noise power $S_0$ has  singularities
at  $eV=n\hbar\Omega$,  when  the  "internal"  frequency  of the
problem $eV/\hbar$ is  a  multiple  of  the  external  frequency
$\Omega$.  We  find that $\partial S_0/\partial V$ is a stepwise
function   of   $V$   that   rises   in   positive   steps    at
$V_n=n\hbar\Omega/e$.  Another  interesting  observation is that
the heights of the steps of $\partial S_0/\partial V$  are  {\it
phase  sensitive},  i.e.,  they  depend  on  the  phase  of  the
transmission  amplitude  in  an   oscillating   way   resembling
A-B effect.
The phase sensitivity of the noise should  be  opposed  to
the  pure  {\rm  dc}  situation  where only the probabilities of
transmission and reflection enter the expression for the  noise,
which  makes  the noise power insensitive to the phase picked by
the wavefunction across the system. In  the  simplest  situation
when the {\rm ac} bias is supplied by alternating flux threading
the  current loop, $\Phi (t)=\Phi_a\sin (\Omega t)$, the heights
of the steps in $\partial S_0/\partial V$  are  proportional  to
the  squares  of  the  Bessel  functions $J_n^2(2\pi\Phi_a/\Phi_0)$,
where $\Phi_0=h c/e$. Let us note that we  are  not  talking
about  the  trivial  effect  of  the  e.m.f.  $-\partial\Phi  /c
\partial t$ induced in the circuit by the alternating flux.  The
effect   in   the   noise will persist in  the  quasistatic  limit
$|\partial\Phi/c\partial t|\ll V$ when the {\rm ac} component of
the current vanishes.

Let us start with recalling general facts about scattering off
an oscillating potential.
We consider a model one dimensional system where electrons are
scattered by alternating scalar and vector potentials $U(x,t)$,
$A(x,t)$ localized in the interval $[-d,d]$, $U(x,t)=A(x,t)=0$
for $|x|>d$. As a function of time they are periodic:
$U(x,t)=\sum_{m=-\infty}^{\infty} U_m(x)\exp(-im\Omega t)$,
where $U_0(x)$ is the static part of the potential, and the other
harmonics $U_m(x)$, $m\ne0$ describe {\rm ac} bias. (Expression for
$A(x,t)$ is similar.) The {\rm dc} bias is expressed in the framework
of the Landauer model as the difference of the population of the
right and the left scattering states. An important difference is
that in our case the states describe {\it inelastic} scattering
because an electron can gain several quanta $\hbar \Omega$ while
passing through the region $[-d,d]$. It will be useful to have
the states expressed through the amplitudes of transmission and
reflection:
    \[{\psi}_{L,k}(x,t)=\left\{ \begin{array}{ll}
e^{-iEt+ikx}
+\sum\limits_n B_{L,n}e^{-iE_nt-ik_nx} & x<-d \\
\sum\limits_n A_{L,n}e^{-iE_nt+ik_nx} &  x>d
  \end{array} \right. ,\]
\begin{equation}{\psi}_{R,k}(x,t)=\left\{ \begin{array}{ll}
\sum\limits_n A_{R,n}e^{-iE_nt-ik_nx} & x<-d \\
e^{-iEt-ikx}+\sum\limits_n B_{R,n}e^{-iE_nt+ik_nx} & x>d
  \end{array} \right. ,
\end{equation}
where   the   amplitudes   $A_{L(R),n},B_{L(R),n}$   are    time
independent.  Here  $E=\hbar^2k^2/2m$, $E_n=E+n\hbar\Omega$, and
$k_n$ are defined by $\hbar^2k_n^2=2mE_n$. The  states  (1)  are
solutions   of   the  Schr{\"o}dinger  equation \\
  \[E\psi (x,t)=[{1\over 2}(-i{\partial\over\partial x }-{e\over
c}A(x,t))^2+U(x,t)]\psi (x,t)\ .\]
   They can be used as a basis to study  transport  through  the
system  the  same  way  it  is  done for the static barrier. The
amplitudes $A_{L(R),n}, B_{L(R),n}$ satisfy unitarity relation,
    \[\sum_{n,n',\alpha,\alpha'} \delta (E_n-E'_{n'})
{\big (} \bar A_{\alpha,n}(E) A_{\alpha',n'}(E')+
\bar B_{\alpha,n}(E) B_{\alpha',n'}(E'){\big )}=
\delta(E-E') \delta _{\alpha \alpha'}\]
that one obtains by the standard reasoning about conservation of
current.

The operator of electric current,
$\hat j(x,t)=-ie\hat \psi^+(x,t)\nabla\hat \psi(x,t)$
is written in terms of second-quantized electrons,
${\hat {\psi}}(x,t)= {\hat {\psi}}_L(x,t)+{\hat {\psi}}_R(x,t)$,
$\hat \psi_L(x,t)= \sum_k \psi_{L,k}(x,t) {\hat a}_k$,
$\hat \psi_R(x,t)=
\sum_k \psi_{R,k}(x,t) {\hat b}_k$,
where  $a_k$   and   $b_k$   are   canonical   Fermi   operators
corresponding  to  the  states (1) coming out of the reservoirs,
the left and the right respectively. It  is  straightforward  to
compute the mean value $I(t)=\langle\hat j(x,t)\rangle$, where
the  brackets  $\langle...\rangle$  stand for averaging with the
density matrix $\rho$ of the reservoirs.  As  usual,  we  assume
absence  of  correlations  in  the  reservoirs, $\hat \rho =\hat
\rho_L \otimes \hat \rho_R$, which physically means  that  after
having  been  scattered  into  a reservoir electrons have enough
time to relax to the equilibrium before they return. Below we
assume   equilibrium   Fermi   distributions    $\rho_{L,R}=
n(E-E_F\pm  eV/2)$.  One  obtains a generalized Landauer formula:
  \begin{equation}
I(t)=\sum_{m=-\infty}^{\infty}  I_m\exp(-im\Omega  t)\ ,
\end{equation}
where it is straightforward to write the coefficients  $I_m$  in
terms    of   the   scattering   amplitudes   $A_{L(R),n}$   and
$B_{L(R),n}$.  Expr.(2)  describes  both  steady   current   and
generation of harmonics in the presence of the {\rm ac} bias.

Quite  similarly  one  can   obtain   an   expression   through
$A_{L(R),n}$,  $B_{L(R),n}$ for the noise. Noise is related with
the correlation function
$S(t_1,t_2)=\langle\!\langle [\hat  j(x,t_1),  \hat  j(x,t_2)]_+
\rangle\!\rangle$.
In the  usual  {\rm dc}  situation  $S(t_1,t_2)=S(t_1-t_2)$  and  its
Fourier transform gives spectral density
$S_\omega=\langle\!\langle j_\omega j_{-\omega}\rangle\!\rangle$
of the noise. In our {\rm ac} case  the  situation  is  somewhat  more
complex because the spectral density $S_\omega$ does not provide
a  complete  description  of  the  noise.  Indeed,  $S(t_1,t_2)$
will depend now separately on $t_1$ and $t_2$,  not  only  on  the
difference $t_1-t_2$. However, it satisfies $S(t_1,t_2)= S(t_1+2\pi/\Omega,
t_2+  2\pi/\Omega)$  resulting  from  the  periodicity of the {\rm ac}
bias. For the Fourier components $\hat j_\omega$ it  means  that
the  average  $\langle\!\langle  j_\omega j_{\omega'} \rangle \!
\rangle$  does  not  vanish  whenever  $\omega+\omega'=m\Omega$,
where $m$ is any integer. Thus, in addition one gets generalized
spectral  densities $S_{\omega,m}=\langle\!\langle j_\omega j_{m
\Omega -\omega} \rangle \! \rangle$, an integer parameter family
of functions. Among them there is the 'ordinary' $m=0$ spectral
density $S_\omega = S_{\omega,0}=\langle\!\langle j_\omega
j_{-\omega} \rangle \! \rangle$, the one easiest to access
experimentally. In what follows we concentrate on it and do not
study other $S_{\omega,m}$, $m\ne 0$.

To compute the noise one has  to  average  the  product  of  two
current  operators  over  the  distribution  in  the reservoirs.
Evaluation of the average is similar to Refs.\cite{2,3,4}, so  we  do
not  need to repeat it here. General expression simplifies quite
substantially in the practically interesting limit of $t_f$,
the  time of flight through the barrier $U(x,t)$ being much shorter
than $2\pi/\Omega$ and $\hbar/eV$.  The  point  is  that  $\hbar
/t_f$  defines  the characteristic scale of energy dependence of
the scattering amplitudes, so the condition $t_f\Omega  \ll  1$,
$t_feV   \ll  \hbar$  enables  one  to  neglect  the  energy
dependence of  $A_{L(R),n}$,  $B_{L(R),n}$  in  the  interesting
energy  domain  $E_F \pm {\rm max}[eV, \hbar \Omega ]$. We also assume
$E_F \gg {\rm max}[eV, \hbar \Omega]$, which allows  to  neglect
the difference of $k_n/m$ and $k/m$, the velocities of scattered
and  incident states, and set $k_n/m=v_F$. It should be remarked
that the  physical  picture  we  discuss  below  is  not  really
dependent  on  any  of  these  assumptions,  they  only make our
expressions more compact. The more  general  case  of  arbitrary
relation  between  $\hbar  /t_f$,  $E_F$,  $eV$,  $\hbar \Omega$
presents no difficulty.

With  the above assumptions made it becomes convenient to use Fourier
transform of the amplitudes $A$ and $B$. Let us define $A_\alpha
(t)=\sum_n  A_{\alpha,n}  \exp  ( -i n \Omega t)$, $\alpha=L,R$.
Similarly we introduce $B_\alpha (t)$, and rewrite Expr.(1) as
    \[{\psi}_{L,k}(x,t)=\left\{ \begin{array}{ll}
e^{ikx}+B_L(t+x/v_F)e^{-ikx} & x<-d \\
A_L(t-x/v_F)e^{ikx} &  x>d
\end{array} \right. ,\]
\begin{equation}{\psi}_{R,k}(x,t)=\left\{ \begin{array}{ll}
A_R(t+x/v_F)e^{-ikx} & x<-d \\
e^{-ikx}+B_R(t-x/v_F)e^{ikx} & x>d
\end{array}\right. .
\end{equation}
(To  obtain  (3)  from (1) we substitute $k_n=k+n\Omega /v_F$ in
the phase shifts $e^{ik_nx}$ and then do the sum over $n$.)  The
amplitudes  $A_{L(R)}(t)$,  $B_{L(R)}(t)$  have clear meaning of
the transmission and reflection amplitudes at given  instant  of
time   for   a  slowly  varying  potential.  The  retarded  time
$t-|x|/v_F$ in Expr.(3) accounts for the finite speed of  motion
after scattering. The unitarity relation now takes the form
  \[|A_{L(R)}(t)|^2+|B_{L(R)}(t)|^2=1,\
\bar A_L(t)B_R(t)+\bar B_L(t) A_R(t)=0\ .\]
  To clarify the character of the simplification thus achieved let
us remark that with Expr.(3) the formula (2) for the current  $I(t)$
becomes just $I(t)=2{e^2\over h}|A(t)|^2eV$ which means that the
current   'adiabatically'   follows   time   variation   of  the
transparency of the barrier according to the  Landauer  formula.
Now  we  shall compute noise and find that, unlike $I(t)$, it is
not reduced to anything trivially related with the static limit.
Let us write the average of two currents
$\langle\!\langle\hat j(t_1)\hat j(t_2)\rangle\!\rangle=$
\[{2e^2\over h^2}\sum_{E,E'}e^{-i(E_k-E_{k'})(t_1-t_2)}
{\big [} |A(t_1)A(t_2)|^2{\big (}n_L(E')(1-n_L(E))+
n_R(E')(1-n_R(E)){\big )}\]
\[+\bar B(t_1)A(t_1) \bar A(t_2)B(t_2) n_R(E') (1-n_L(E))+
\bar A(t_1)B(t_1) \bar B(t_2)A(t_2) n_L(E') (1-n_R(E)){\big ]} .\]
To compute $S_\omega$ we have to do Fourier transform and substitute
Fermi distributions \break $n_{L(R)}(E)$.
Explicit calculation yields the result
\begin{equation}S_{\omega}=
{2e^2\over \pi} \sum\limits_n 2 N_0(\omega-n\Omega)
|(|A|^2)_n|^2 + N_1(\omega ,n\Omega+eV) |(A\bar B)_n|^2, \end{equation}
where \[ N_0(x)=\int (n(E-x)+n(E+x))(1-n(E))dE= x{\rm coth}(x/2T)\ ,\]
\[ N_1(x,y)=N_0(x+y)+N_0(x-y)={x \sinh (x/2T) - y \sinh (y/2T)\over
\cosh (x/2T) - \cosh (y/2T)}\ ,\]
and $(...)_n$ denotes Fourier components, e.g., $(A\bar B)_n={\Omega
\over 2\pi}\int A(t)\bar B(t) e^{in\Omega t}dt$. Expr.(4)
describes the noise as function of $eV,\ \Omega,\ \omega$ and $T$.
The behavior is simplest at $T=0$ when $N_0(x)=|x|$, $N_1(x,y)=|x+y|+|x-y|$.
Given by Expr.(4) as a weighted sum of terms
like $|n\Omega + eV \pm\omega|$, $|\omega-n\Omega|$ the noise
$S_{\omega}$ will then depend on $V,\ \Omega,\ \omega$ in a
piecewise linear way, changing from one slope to another when
$n\Omega+eV\pm\omega$ or $\omega-n\Omega$ equals $0$. This
condition defines the locations where $S_{\omega}$ has
singularities. They are cusps, sharp at $T=0$ and rounded on the
scale $T$ at $T>0$.

With  the  general  Expr.(4) one  can  explore  the noise in all
possible  limiting situations that  one   obtains   for   different
combinations of $eV,\ \Omega,\ \omega$ and $T$. Particularly
interesting  for  us  will be the case $T=0$, $\omega =0$
corresponding to the noise
$S_0 =\langle\!\langle j_\omega j_{-\omega}\rangle\!
\rangle_{\omega \rightarrow 0}$ measured at low  frequency.  Let
us  remark  here  that setting $\omega =0$ means only that $\omega$ is
small compared to the parameters $eV$ and $\Omega$  that  define
the  width of the frequency band of the excess noise. Such $\omega$ may
still be much higher than the band width for  other  sources  of
noise,  e.g., the $1/f$. Let us concentrate on the dependence of
$S_0$ on $V$. It is a piecewise linear function which is easiest
to characterize by its derivative,
\begin{equation}\partial S_0/\partial V={2e^3\over \pi }\sum\limits_n \lambda_n
\theta (eV-n\hbar\Omega),  \end{equation}
where $\lambda_n=|(\bar A_L B_R)_n|^2$ and  $\theta  (x)=1$  for
$x>0$,  $-1$  otherwise.  The function $\partial S_0/\partial V$
rises in positive steps  at  all  $V_n=\hbar  \Omega  n/e$  (see
Fig.~\ref{figure1}),  the  property  that  can be alternatively
formulated as convexity of $S_0(V)$ in $V$.

The  meaning  of  the  singularities  in  $S_0(V)$ was clarified
recently in a study of  the  statistics  of  transmitted  charge
\cite{IL}.  The  generating  function of the charge distribution
was expressed through the single-particle scattering matrix, and
it  was found  that  the distribution  arises   from   Bernoulli
statistics (i.e., it is a generalized binomial distribution).   The
frequencies of attempts were  given  as  function  of  $V$  and
$\Omega$. The probabilities of outcomes of a single attempt were
found  in  terms  of many-particle scattering amplitudes, and it
was   shown    that    they    change    at    the    thresholds
$V_n=n\hbar\Omega/e$ in a discontinuous way due  to  statistical
correlation  in  the  outgoing  channels  of the scattering. The
discontinuity manifests itself  in  the  second  moment  of  the
distribution  that  corresponds  to the noise $S_0(V)$ discussed
above.

There is  an  interesting  and  simple  example  where  one  can
explicitly  evaluate the heights of the steps. Let us consider a
junction with ideal leads bent into a loop of  length  $L$  (see
inset of Fig.~\ref{figure1})  and  placed  into  an  external
magnetic field varying with
time. In this  problem  the  junction  is  the  only  source  of
scattering. For simplicity let us assume  that only one scattering
channel is involved and that
the junction is symmetric,  $A_L=A_R=A$,  $B_L=B_R=B$.  The  {\rm ac}
bias is supplied by the alternating flux of the magnetic field
through the loop, $\Phi (t)=\Phi_a\sin (\Omega t)$. Also let us suppose
that the magnetic field is quasistatic, i.e., the time of
flight through the system, $t_f=L/v_F$ is much shorter than
$2\pi/\Omega$, that makes it possible to introduce the time
dependent amplitudes $A_{L(R)}(t),\ B_{L(R)}(t)$ as it was
discussed above. As is common, in such a situation the vector
potential can be treated semiclassically, and one can write the
wavefunction as $\psi (x,t)= \exp ({ie\over\hbar
c}\int_{-\infty}^{x} A(x')dx') \psi_0 (x,t)$, where $x$ is the
coordinate along the lead and $\psi_0 (x,t)$ is found by solving the
Schr{\"o}dinger equation in the absence of the magnetic field.
Thus all the dependence on the magnetic flux can be accumulated
in the phase of the transmission amplitude,
   \[A_{R(L)}(t)=\exp (\pm i\Phi (t)/\Phi_0) A$, $B_{R(L)}(t)=B, \]
where $\Phi_0=h c/e$ is single electron flux quantum.  Since
$|A(t)|^2=D=const$    the    current    is   time   independent:
$I={2e^2\over  h}DV$.  According  to  Expr.(4)  $S_{\omega}$  is
written through the Fourier components of $\bar  A_L(t)  B_R(t)$  in
this    case    given    by the  Bessel   functions:   $(\bar   A_L
B_R)_n=J_n(2\pi\Phi_a/\Phi_0)AB$. Thus we find
     \begin{equation}S_{\omega}={2e^2\over   \pi}    {\big    [}
2N_0(\omega     )D^2+     \sum\limits_n     N_1(\omega     ,n\Omega+eV)
D(1-D)J_n^2(2\pi\Phi_a/\Phi_0){\big ]}. \end{equation}
    The  heights  $\lambda_n$  of   the   steps   in   $\partial
S_0/\partial  V$  are  then  given by
  \begin{equation}\lambda_n=
D(1-D) J_n^2(2\pi\Phi_a/\Phi_0). \end{equation}
  They oscillate as function of $\Phi_a / \Phi_0$ and vanish  at
the nodes of Bessel functions.

Exprs.(6),(7)  illustrate one important feature of the noise
in the {\rm ac} biased system, the sensitivity to the {\it phase} of
the transmission amplitude $A$. By varying the amplitude
$\Phi_a$ of the alternating flux one can make $\lambda_n$ vanish
separately for each harmonic $n\Omega$ of the {\rm ac} frequency. This
should be compared with the case of the {\rm dc} bias where the
noise is expressed only through $|A|^2$ and thus cannot be phase
dependent. We call the oscillating dependence (6),(7) {\it
non-stationary Aharonov-Bohm effect}. To compare it with the
usual A-B effect let us recall that the latter is observed
as an oscillation of the {\rm dc} conductance under
variation of flux in the situation
when one has interference of transmission
amplitudes corresponding to different classical trajectories of
a quantum system, e.g., in a conductor with multiply connected
leads forming one or several closed loops. The {\rm dc} A-B effect
cannot be observed in the single path geometry like Fig.1.
Alternatively, the non-stationary A-B effect appears as a result
of interference of the right and left scattering states
travelling in the  opposite directions along same path and having
energies shifted by $n\Omega$. It is clear from our discussion that
such interference does not contribute to the {\rm ac} conductance but
is important for the noise and, therefore, one obtains the
non-stationary A-B effect in the noise even in the topologically
trivial situation of Fig.1.

One can derive a sum rule:
\begin{equation}\sum\limits_n \lambda_n={\Omega\over 2\pi}
\int_0^{2\pi/\Omega} D(t)(1-D(t))dt \ ,
\end{equation}
where $D(t)=|A(t)|^2$. For $\lambda_n$ given by Expr.(7) it
follows from the definition of the Bessel functions. In the general
case of Expr.(5) the sum rule is obtained by applying
Plancherel's formula to Fourier components of $A(t)\bar B(t)$.
The sum rule clarifies the relation of our problem
with the previous calculation\cite{2,3,4} of the noise
in the pure {\rm dc} case
for which the result does not depend on the phase of $A(t)$.
When the limit is taken $\Omega\rightarrow 0$, $V=const$, the
steps in $\partial S_0/\partial V$ do not vanish
but  just  move closer to zero, thus effectively condensing then
all together in a single step at $V=0$. The height of this  step
is not phase sensitive and is simply given by the expression (8)
for the {\rm dc} noise averaged over the period $2\pi/\Omega$.

It is worth mentioning that our results for $S_\omega$ are quite
general. Indeed, it is clear after what have been said that the
singularities at $V=n\hbar\Omega/e$ are only due to the sharp
edge of the Fermi distribution, and not related with any
specific geometry assumed for the junction. Because of that the
phenomenon should be displayed by any coherent conductor,
provided that the main source of inelastic scattering is the {\rm ac}
potential.  The  reason  is that an elastic scattering, if any, can smear
the Fermi distribution of momenta but it  will  not  affect  the
sharpnes of the step in the energies
distribution, and our effect is sensitive only to the latter.
The same remark applies to the oscillating dependence  of
the singularities on the amplitude of the {\rm ac} signal.

Let us briefly discuss a generalization of the system shown in
Fig.1 where the loop is not an ideal lead but a real metallic
wire with disorder, i.e., instead of one scatterer there are now
many of them uniformly distributed over the bulk of the wire.
Most interesting is the case of a purely coherent conductor for
which the energy relaxation time $\tau_E$ and the phase breaking
time $\tau_\phi$ are much longer than the flight time $t_f$. (One
can estimate $t_f\approx \hbar/E_c$, where $E_c$ is Thouless'
energy $\hbar D/L^2$.) In such a system transport is described
by channels of the scattering matrix with transmission
coefficients $T_m$ assigned to each channel\cite{7}.
In the {\rm dc} case the noise can be written\cite{4} in terms of
$T_m$ as $S_0={2e^2\over \pi} \sum_m T_m(1-T_m)eV$. In the
presence of the alternating flux the extension of our formalism
can be carried out easily and one obtains expressions similar to
(6) and (7), with $D^2$ and $D(1-D)$ replaced by $\sum_m T_m^2$
and $\sum_m T_m (1-T_m)$ respectively.
However, the limitations under which the result is valid,
$eV\ll\hbar/t_f$, $\Omega\ll 1/t_f$, are now slightly more
stringent than for Exprs.(6),(7) because the flight time $t_f$
is longer.

A more fundamental limitation to the general validity of our
calculation is in the assumption that the flux threads only the
phase coherent part of the conductor. It would certainly be of
interest  to  better understand the opposite limit when the {\rm
ac} voltage increases smoothly over a distance  much  larger
than the phase breaking length $L_\phi=\sqrt {\tau_\phi/D}$.

To summarize, we studied current and noise in a conductor driven
by {\rm dc} and {\rm ac} and we expressed them through time-dependent one
particle scattering amplitudes. In the quasistatic limit of
short time of flight through the conductor the current is given
by the Landauer formula with time-dependent transmission
coefficient, i.e., by a trivial generalization of the static
case. The situation with the noise is quite different because of
the two-particle interference. Spectral density of the noise
$S_\omega$ depends on the scattering amplitudes in such a way
that the phases do not drop out, and this leads to a
non-stationary Aharonov-Bohm effect. Because of the way the Fermi
statistics affects the two-particle interference the noise
measured at $T=0$ is singular at $\omega =\pm eV/\hbar
+m\Omega$, where $m$ is any integer. To illustrate the phase
sensitivity of the noise we consider a conducting metallic loop
in which the {\rm ac} signal is supplied by an oscillating magnetic flux.
Because of the sensitivity to the phase of transmission
amplitude the strengths of the singularities in the noise
display oscillatory dependence on the amplitude of the {\rm ac} flux
given by squares of the Bessel functions.

\acknowledgements
We are indebted to D. E. Khmelnitskii for drawing our attention
to the problem of harmonic generation in coherent conductors,
and to J. Hajdu for illuminating discussions.\\
Research of L. L. is partly supported by Alfred Sloan fellowship.
The work of G. L. is performed within the research program of the
Sonderforshungsbereich 341, K\"oln-Aachen-J\"ulich.

\begin{figure}
\caption{ Differential noise $\partial S_0/\partial V$ at $T=0$
given by Exprs.(5),(7) is plotted against V for three flux amplitudes:
(1) $\Phi_a=5\Phi_0/4\pi$;
(2) $\Phi_a=7\Phi_0/2\pi$;
(3) $\Phi_a=23\Phi_0/4\pi$.
({\it Inset:} Junction with leads bent in a loop through which alternating
magnetic flux is applied.)}
\label{figure1}
\end{figure}

\end{document}